\documentclass[runningheads]{llncs}
\usepackage[T1]{fontenc}
\usepackage{graphicx}
\usepackage{balance} 
\usepackage[svgnames]{xcolor}
\usepackage{tikz}
\usetikzlibrary{arrows.meta, positioning}
\usepackage{xspace}
\usepackage{array}
\usepackage{listings}
\usepackage{csquotes}
\usepackage{footmisc}

\usepackage{booktabs}
\usepackage{amssymb}
\usepackage{wrapfig}

\usepackage{hyperref}
\usepackage{xurl} 
\usepackage{url}
\usepackage[linesnumbered,ruled,vlined]{algorithm2e}

\graphicspath{{Figures}}

\lstdefinelanguage{EmfaticText}{
  morekeywords={@namespace,package,class,attr,val,ref,extends},
  keywordstyle=\color{MediumBlue}\bfseries,
  morekeywords=[2]{String,int},
  keywordstyle=[2]\color{Green},
  sensitive=true,
  morecomment=[l]{//},
  morestring=[b]",
  alsoletter={*},
  basicstyle=\ttfamily\small,
  stringstyle=\color{MidnightBlue},
  commentstyle=\color{gray}\itshape,
  columns=flexible,
  keepspaces=true,
  showstringspaces=false,
  tabsize=2,
  frame=lines,
  numberstyle=\scriptsize,
  captionpos=b
}

\newcommand*{\classname}[1]{\emph{#1}}

\newcommand{\reqdiscover}{R1\xspace}
\newcommand{\reqdiscoverdesc}{Agent discoverability}
\newcommand{\reqmultipart}{R2\xspace}
\newcommand{\reqmultipartdesc}{Multi-part messaging}
\newcommand{\reqmultiturn}{R3\xspace}
\newcommand{\reqmultiturndesc}{Multi-turn conversations}
\newcommand{\reqasync}{R4\xspace}
\newcommand{\reqasyncdesc}{Asynchronous and streaming communication}
\newcommand{\reqobserve}{R5\xspace}
\newcommand{\reqobservedesc}{Agent observability}
\newcommand{\reqinterop}{R6\xspace}
\newcommand{\reqinteropdesc}{Interoperability}
\newcommand{\reqaccess}{R7\xspace}
\newcommand{\reqaccessdesc}{Access control}

\newcommand{\BibTeX}{\rm B\kern-.05em{\sc i\kern-.025em b}\kern-.08em\TeX}

\newif\ifsubmissionanon
\submissionanonfalse   

\newcommand{\ifanon}[2]{%
  \ifsubmissionanon #1\else #2\fi
}

\begin{document}

\title{A Comparative Study of MCP and A2A for Inter-Agent Coordination in LLM-Based Systems}

\titlerunning{A Comparative Study of MCP and A2A for Inter-Agent Coordination}
%
\author{Ionut Predoaia\inst{1}\orcidID{0000-0002-2009-4054} 
\and Tuong Manh Vu\inst{1}\orcidID{0000-0002-2540-8825} \and
Konstantinos Barmpis\inst{1}\orcidID{0000-0002-0864-0956}
\and
\\ Dimitris Kolovos\inst{1}\orcidID{0000-0002-1724-6563}
\and
\\ Antonio García-Domínguez\inst{1}\orcidID{0000-0002-4744-9150}}
\authorrunning{I. Predoaia et al.}
%
\institute{University of York, York, United Kingdom \\ \email{\{ionut.predoaia, tuong.vu, konstantinos.barmpis,\\ dimitris.kolovos, a.garcia-dominguez\}@york.ac.uk}}
\maketitle              
\begin{abstract}
Recent industry practice has seen the rapid emergence of agentic systems composed of heterogeneous, tool- and LLM-mediated agent components, raising practical questions about inter-agent coordination and protocol design. This paper presents an implementation-grounded comparison of the Model Context Protocol (MCP) and the Agent2Agent (A2A) protocol, from a multi-agent systems engineering perspective, using an inter-agent coordination scenario involving LLM-based agents. We evaluate an MCP-based and an A2A-based multi-agent implementation of the same software engineering task against a set of requirements derived from prior literature and discussions with industry partners, including agent discoverability, multi-part messaging, multi-turn conversations, asynchronous communication, observability, interoperability, and access control. The results evidence that MCP can support inter-agent coordination in constrained LLM-based systems through a comparatively lightweight implementation model with lower coordination complexity, although coordination concerns such as conversational state management and task lifecycle handling must be implemented explicitly at the application layer. In contrast, A2A provides richer native support for stateful, multi-turn coordination through protocol-level abstractions for tasks and lifecycle management, but this comes with substantially greater implementation and coordination complexity. Given the narrow scope of the evaluated coordination pattern, these findings are presented as design observations from an empirical experience report rather than general claims of protocol suitability or superiority across broader classes of MAS, highlighting trade-offs and how protocol abstractions shape the distribution of coordination responsibilities in contemporary agentic systems.

\keywords{Multi-Agent Systems \and Large Language Models \and Inter-Agent Coordination \and Model Context Protocol \and Agent2Agent Protocol \and Model Generation \and Experience Report \and Implementation Study}
\end{abstract}

\section{Introduction}

Recent advances in large language models (LLMs) have led to the emergence of contemporary agentic systems in which multiple, specialised agent components collaborate to accomplish complex tasks. In the classical Multi-Agent Systems (MAS) literature, Wooldridge defines agents under the weak notion of agency as autonomous computational entities that are situated in an environment, perceive and act upon that environment, exhibit reactive and proactive behaviour in pursuit of their objectives, and are capable of interacting with other agents through explicit communication mechanisms~\cite{wooldridge1995intelligent}. This weak notion contrasts with stronger notions of agency, which assume rich internal mental states~\cite{shoham1993agent} or normative social semantics~\cite{singh2000social}. Nevertheless, in contemporary LLM-based practice, agents are often defined more pragmatically as LLM-enabled components that participate in multi-step task execution within a software system. This usage is closely related to the LangChain notion of agents~\cite{langchain_agent} as LLM-driven components within an application, which frames agentic behaviour as a spectrum based on the extent to which an LLM determines system control flow. 

Inspired by the LangChain perspective on agentic systems, this paper adopts a pragmatic notion of agents as LLM-enabled components capable of executing multi-step tasks within a software system. The paper's focus is on coordination patterns in contemporary agentic systems, where heterogeneous agents interact through tools, APIs, and communication protocols. Although these systems differ architecturally from many classical MAS settings, they raise related engineering concerns, including agent discoverability, interaction structure, and task lifecycle coordination.

The paradigm of decomposing complex tasks into collaborations among multiple, specialised agents introduces substantial integration challenges, requiring careful consideration of communication protocols and interoperability standards. From 2024 onwards, there has been a significant industry push toward inter-agent collaboration, accompanied by the proposal of several agent communication protocols, with recent surveys comparing their relative strengths and weaknesses~\cite{yang_survey_2025,ehtesham_survey_2025}. Among these protocols, two have gained the most traction within the developer community, as reflected by GitHub activity (in terms of repository stars) and reported adoption: the Model Context Protocol (MCP)~\cite{mcp-web} and the Agent2Agent (A2A) protocol~\cite{a2a-github}. These two protocols differ fundamentally in their original design intent: MCP was designed to streamline the connection of LLM-based applications to tools, data sources, and workflows, whereas A2A was specifically designed to support inter-agent communication and collaboration across heterogeneous agents.

Prior academic work comparing agent interoperability protocols characterises MCP and A2A as addressing different layers of agentic systems, with MCP primarily supporting structured access to tools and data sources, and A2A focusing on explicit agent-to-agent communication and task delegation~\cite{yang_survey_2025,ehtesham_survey_2025}. Several studies argue for a complementary use of MCP and A2A rather than treating them as competing alternatives~\cite{li2025glue,liao_agentmaster_2025,tupe_demonstrating_2025,ghosh2025agentic,jeong2025practical,jeong_study_2025}. Application-oriented works that employ both protocols consistently follow this layered architecture across domains such as incident response, engineering design, digital twins, and automation~\cite{tupe_demonstrating_2025,ghosh2025agentic,jeong2025practical,gholizadeh2025agent,dinh2025llm,ghosh2025beyond,habler_building_2025}. 

Practitioners have explored the use of MCP beyond its original design scope, notably as an infrastructure for supporting coordination in agentic systems. This exploration is reflected in practitioner discussions, industry presentations~\cite{voss_mcp_2025,colvin_mcp_2025}, and technical blog posts~\cite{aldridge_open_2025,dibia_can_2025}, where MCP is often positioned as a unifying integration layer for orchestrating agent-based components. We treat these practitioner-oriented sources not as scientific evidence, but as indicators of emerging engineering practices in rapidly evolving LLM-based agentic systems. Collectively, they suggest a growing tendency to repurpose general-purpose integration protocols, such as MCP, to address coordination concerns beyond their original design scope. To the best of our knowledge, the existing literature has neither investigated the use of MCP for enabling inter-agent coordination nor provided an empirical comparison between MCP and A2A in this setting. This motivates the implementation-grounded comparative study presented in this paper.

MCP is increasingly adopted as a standard mechanism for providing tool access to LLM-based systems, and practitioners may therefore already rely on MCP as part of their existing systems. In the context of LLM-based systems that already rely on MCP for tool integration, the introduction of multi-agent coordination raises a practical design question: whether MCP can be reused to support inter-agent communication, or whether a protocol explicitly designed for agent-to-agent communication, such as A2A, is required. This question has been actively debated in industry discussions, where MCP is often argued to already satisfy many practical requirements for coordinating agent-based components.

The software engineering use case that we have selected for the comparison in this paper is intentionally narrow in scope and instantiates a minimal yet representative coordination pattern involving task delegation, iterative refinement with feedback, and validation. This coordination pattern recurs across many contemporary agentic systems (e.g., iterative cycles of planning, execution, and verification), making it suitable for protocol-level analysis. To study the practical use of MCP and A2A in this setting, we implement a multi-agent collaboration that solves a common software engineering task: given a natural language prompt, generate an object-oriented domain model.


This paper contributes an implementation-grounded, empirical engineering study comparing MCP and A2A by applying both protocols to the same multi-agent coordination scenario. Our objective is to characterise how differing protocol abstractions shape the distribution of coordination responsibilities between protocol mechanisms and application-specific orchestration logic in a concrete collaboration pattern. Furthermore, we evaluate the MCP-based and the A2A-based implementations against a set of requirements derived from prior literature and discussions with industry partners, including agent discoverability, multi-part messaging, multi-turn conversations, asynchronous communication, observability, interoperability, and access control. This work is intended as an empirical experience report on applying emerging industry protocols in a concrete LLM-based MAS, rather than as a benchmark or performance evaluation.

The results indicate that MCP can support inter-agent coordination in constrained LLM-based systems through a comparatively lightweight implementation model with lower coordination complexity, fewer protocol primitives, and reduced orchestration overhead, although key coordination concerns such as conversational state management and task lifecycle handling must be implemented explicitly at the application layer. In contrast, A2A provides richer native support for stateful, multi-turn coordination through protocol-level abstractions for tasks and lifecycle management, but this entails substantially greater implementation and coordination complexity. Given the narrow scope of the evaluated coordination pattern, these findings are intended as design observations rather than general claims of protocol suitability across broader classes of MAS.

\section{Background}

\subsection{Agentic Communication}

In classical MAS, interaction is usually discussed in relation to an \emph{environment} that mediates perception and action, and that can be treated as a first-class abstraction to clarify what responsibilities belong to agents versus shared infrastructure~\cite{weyns_environment_2007}. A closely related engineering perspective separates \emph{agents} (decision-making entities) from the \emph{artifacts} they use. Artifacts provide controllable, explicit capabilities in the working environment and can be either \emph{resources} (for accessing and sharing information) or \emph{tools} (for performing actions)~\cite{ricci_artifacts_2007}. This abstraction highlights that ``communication'' is not just message exchange: it is part of the broader machinery by which agents coordinate action and share responsibility.

Historically, agent communication has been supported by standardised message languages that expose the intent of messages (e.g., \emph{ask}, \emph{tell}, \emph{reply}), with canonical examples including KQML (Knowledge Query and Manipulation Language)~\cite{finin_kqml_1994} and the FIPA-ACL specification~\cite{fipa_acl_2002}. These early formal models rely on a limited set of standardised interaction patterns and assume \emph{mentalist} semantics, i.e., they define communicative acts in terms of an agent's internal beliefs and intentions. For example, when an agent sends a message (such as \emph{tell}), it is attempting to make the receiving agent believe something (i.e., changing its mental state). Such standards are often criticized as \emph{outdated}~\cite{chopra_toolsuite_2025} for a system of heterogeneous and autonomous agents, because agents' internal states are hidden. Because of these limitations, many modern approaches prefer \emph{social} semantics, in which the meaning of communication is defined in terms of \emph{publicly observable} interaction state (e.g., commitments created, delegated, or discharged) and the explicit \emph{protocol} being enacted, rather than in terms of beliefs and intentions. This shift supports reasoning and verification over interactions between heterogeneous and autonomous agents, precisely because it does not require access to, or agreement on, any particular internal mental representation~\cite{christie_kiko_2023,singh_langshaw_2024,chopra_azorus_2025,chopra_toolsuite_2025}. These approaches treat interaction protocols as first-class engineering artifacts and emphasise declarative protocol specifications that can be enacted and reasoned about independently of any single agent architecture.

In contemporary LLM-based agentic systems, agents can collaborate with other agents, decompose tasks, call external tools (APIs, databases, code execution), and exchange intermediate artifacts. As a result, agent orchestration and tool access become important engineering concerns. This distinction between agents and tools also has direct implications for communication protocol design: agent-to-tool protocols typically emphasise structured invocation and resource access, whereas agent-to-agent protocols must support richer interaction patterns including negotiation, delegation, and multi-turn coordination.

Recent surveys of agent interoperability protocols highlight that emerging proposals often target different aspects of an agentic system (e.g., context/tool vs. inter-agent)~\cite{yang_survey_2025,ehtesham_survey_2025}. From an engineering standpoint, this leads to a recurring design decision: whether a protocol primarily standardises \emph{agent-to-tool} integration (structured invocation of external capabilities) or \emph{agent-to-agent} interaction (turn-taking, delegation, long-running tasks, and progress reporting). To make this distinction concrete, the next two subsections introduce MCP and A2A as representative protocols: one aimed primarily at tool integration, the other at inter-agent collaboration.

\subsection{Model Context Protocol (MCP)}

The Model Context Protocol (MCP)~\cite{mcp-web} is a JSON-RPC client–server interface for secure context ingestion and
structured tool invocation which streamlines the integration of LLMs with external data sources and tools. To this end, it supports flexible plug-and-play tools, safe infrastructure integration, and compatibility across LLM vendors \cite{ehtesham_survey_2025}. The usage of MCP is characterised by the presence of three architectural components that structure how LLM-based agents interact with external resources and tools. In this architecture, the \emph{host} refers to agents responsible for interacting with users, understanding and reasoning through user queries, and selecting tools. The \emph{client} is connected to a host and is responsible for providing descriptions of available resources, whereas the \emph{server} is connected to the resource(s) and establishes a one-to-one connection with the client, providing the required context from the resource(s) to the client.

MCP primitives constitute the core protocol abstractions and define the types of contextual information and executable actions that clients and servers can exchange. They specify both \emph{what} information can be shared with LLM-based applications and \emph{which} actions can be performed. MCP defines three first-class primitives that servers can expose: \emph{resources}, \emph{tools}, and \emph{prompts}. \emph{Resources} are data sources such as file contents, database records, or API responses; \emph{tools} are executable functions such as file operations, API calls, or database queries; and \emph{prompts} are reusable templates that structure interactions with language models.

\subsection{Agent2Agent (A2A) Protocol }
\label{sec:a2a-protocol}

A2A is a lightweight, web-native protocol for enterprise-oriented agent communication, building on existing standards (i.e., HTTP and JSON-RPC) to support asynchronous, long-running interactions across heterogeneous agents. It enables modality-independent message exchange while preserving opaque execution, allowing agents to collaborate without exposing internal logic or proprietary capabilities. The official specification of A2A~\cite{a2a_specification} is expressed as a set of JSON-RPC messages to be exchanged between an A2A client and an A2A server.
An A2A server hosts one or more \emph{agents}, which are described using \emph{agent cards} served from the same server, or from a registry or catalogue.
Agent cards are JSON documents served over HTTP that provide structured metadata advertising each agent's capabilities and providing the necessary communication endpoints.

A2A defines a specific structure for a message, which can be divided into multiple parts.
Within a specific turn of communication between a client and an agent, the client will send a message with its request to the agent, who can reply directly with another message (in case a response can be immediately given), or with a \emph{task}.
A task represents a stateful and long-running piece of work whose progress needs to be tracked by the client and the server.

Tasks have a well-defined lifecycle within A2A. A task is created in response to a message from the client when the agent considers that producing the response will take some time, is immediately returned to the client with a unique ID for later reference, and may provide interim updates on changes in the \emph{artifacts} produced by the task or in the \emph{state} of the task. At some point, the task will reach a terminal state (e.g., \emph{completed}, \emph{cancelled}, \emph{failed}, or \emph{rejected}), or a paused state (\emph{input-required} or \emph{auth-required}), allowing the client to provide additional information for the agent to resume execution.


\section{Methodology}
\label{sec:methodology}

\subsection{Software Engineering Task}
\label{sec:task}

To study the practical use of MCP and A2A in a realistic setting, we implement and evaluate a multi-agent collaboration that solves a common software engineering task: given a prompt in natural language, produce an object-oriented model of the domain. This task instantiates a coordination pattern in which a solution-producing component is embedded within an iterative validation and refinement loop involving specialised supervisory components.

The object-oriented model is expressed in a machine-readable format, such that it can be used by specialised tools to automatically generate code and documentation. While formats such as PlantUML~\cite{plantuml_website} and MermaidJS~\cite{mermaid_website} may be sufficiently popular to be directly generated by existing foundational models (e.g., GPT-4), these formats are primarily intended for visualisation and not for automated model processing (e.g., generating Java code or HTML documentation of a domain model).
Instead, domain models are produced in the Emfatic~\cite{emfatic_website} textual language of the Eclipse Modelling Framework (EMF)~\cite{steinberg_emf_2008}, from which it is possible to generate Java code and perform other automated transformations.

We propose the multi-agent collaboration defined in Algorithm~\ref{alg:collaboration-workflow} to solve the domain model generation task described above. The collaboration structure is fixed across both protocol implementations in order to isolate the effects of protocol-level abstractions on coordination and interaction design. The algorithm takes as input a prompt describing a domain, and a maximum number of retries, and then returns either a validated artifact or a failure outcome if no valid artifact can be generated within the allowed number of attempts.

\SetKw{Continue}{continue}

\begin{algorithm}[t]
\caption{Collaboration Agent: Workflow for Model Generation}
\label{alg:collaboration-workflow}

\KwIn{User prompt $P$, maximum retries $N$}
\KwOut{Validated artifact $M$ or failure}

$attempts \gets 0$, 
$validated \gets false$\;

\While{$attempts < N$ and not $validated$}{

    $attempts \gets attempts + 1$\;

    $M \gets \textsc{SolutionAgent.Generate}(P)$\;

    $syntaxFeedback \gets
    \textsc{SyntacticSupervisorAgent.Validate}(M)$\;

    \If{$syntaxFeedback.status = invalid$}{
        $P \gets \textsc{RefinePrompt}(P, syntaxFeedback)$\;
                
        \Continue\;
    }

    $semanticFeedback \gets
    \textsc{SemanticSupervisorAgent.Validate}(P, M)$\;

    \If{$semanticFeedback.status = invalid$}{
        $P \gets \textsc{RefinePrompt}(P, semanticFeedback)$\;
        
        \Continue\;
    }

    $validated \gets true$\;
}

\Return{$M$ if $validated$ else failure}\;

\end{algorithm}

\begin{enumerate}
\item A \emph{collaboration agent} orchestrates the entire agentic workflow described in Algorithm~\ref{alg:collaboration-workflow}. It receives the description of the domain from the user, and forwards it as a new request to the \emph{solution agent} (line 4). As the focus of this paper is to compare protocol-level support for coordination, the internal implementation strategy of individual agents is not treated as a variable in the analysis. Accordingly, the workflow used in this paper is fixed; computing collaborations on the fly is outside the scope of this work.

\item A \emph{solution agent} uses an LLM to drive the production of the domain model, which it returns to the collaboration agent. This LLM may directly generate the domain model in the Emfatic language, or it may use tools to build up a model and then generate Emfatic source code, by following the ReAct architecture~\cite{yao_react_2023}. The choice will depend on the capabilities and training data used by the selected LLM: larger models may be able to directly generate Emfatic, whereas smaller models may require the assistance of purpose-specific tools and the ReAct architecture. The possibility that the solution agent produces invalid, incomplete, or no output at all is explicitly considered in order to exercise iterative interaction and feedback in the collaboration (up to a certain number of attempts), with failures being signalled by the solution agent through detailed error messages to support fixing the generated model.

The solution agent may produce output that does not conform to the Emfatic grammar or semantics. For that reason, the collaboration agent will take the proposed solution and forward it to the \emph{syntactic supervision agent} (line 5), which will parse the solution using an Emfatic parser.
\begin{itemize}
\item If any errors are reported (lines 6--8), the syntactic supervision agent will inform the collaboration agent that the solution is invalid, and include feedback with the specific errors. The collaboration agent will then pass on the feedback to the solution agent, who will make another attempt to produce a valid solution (up to a certain number of attempts).

\item If no errors are reported, the agent will only indicate that the solution is valid, and the collaboration agent will proceed to the next step.
\end{itemize}

\item Although syntactically correct, the output may not meet all the requirements of the original description. For that reason, the collaboration agent will pass on the original description of the domain and the generated Emfatic source to a \emph{semantic supervision agent} (line 9). This agent will use an LLM that has been instructed to read Emfatic code through \emph{few-shot prompting}, by providing it a few examples of Emfatic syntax within the prompt.
\begin{itemize}
\item If any issues are detected (lines 10--12), the semantic supervision agent informs the collaboration agent, who forwards the feedback to the solution agent and asks for an updated solution.
\item If the semantic supervision agent finds no issues in the solution, the collaboration agent returns the final solution (lines 13 and 14).
\end{itemize}
\end{enumerate}

\subsection{Requirements}
\label{sec:requirements-evaluation}

To structure the analysis of the protocol-based implementations of the collaboration scenario described in Section~\ref{sec:task}, we derive a set of requirements informed by recent survey literature~\cite{yang_survey_2025,ehtesham_survey_2025} and by practical needs articulated by industry partners within the \ifanon{[ANONYMISED]}{MOSAICO} research project \ifanon{}{\cite{MOSAICO}}. MOSAICO investigates collaborative, LLM-based agent ecosystems for software development, focusing on the design of communication protocols, intelligent agents spanning the software development lifecycle, and practical conditions for their effective and responsible adoption in real-world settings. 
These requirements are intended as analytically useful dimensions that reflect recurring concerns in the design and engineering of coordinated agent systems, rather than as an exhaustive or normative evaluation framework. In particular, they are used to structure a qualitative, implementation-grounded comparison of protocol characteristics, rather than to support quantitative benchmarking or claims of protocol superiority. Specifically, \ifanon{}{within MOSAICO, }functional and non-functional requirements were elicited from practical use cases provided by \ifanon{industry partners from the domains of finance, manufacturing, software, and hardware sectors (Note: Partner names and specific industries omitted for double-blind review)}{business partners in diverse industries, including banking (National Bank of Greece~\cite{bank_greece}), aerospace (Collins Aerospace~\cite{collins_aerospace}), immersive technologies (Immersion~\cite{immersion_website}), and IoT (UNPARALLEL~\cite{unparallel_website})}.

\begin{description}
\item[\reqdiscover: \reqdiscoverdesc] The protocol should have a way to list the collaboration / solution / supervision agents implemented by a given \emph{agent server}, and describe their most important features.
The description should include: identification (name and description), skills, message formats of input and output, and security formats (authentication schemes and authority).

\item[\reqmultipart: \reqmultipartdesc] The protocol should separate the natural language parts of their responses from their structured parts (e.g., valid/invalid evaluations, or the generated Emfatic source code), resulting in \emph{multi-part} messages.

\item[\reqmultiturn: \reqmultiturndesc] The protocol should allow the solution agent to request further clarification on the domain description if it considered necessary.
Note that for this experiment, the implemented agents did not perform requests for clarification: instead, we reviewed the protocol specifications for the support of such a mechanism.

\item[\reqasync: \reqasyncdesc] A user should be able to either ask regularly for updates on the generation of the solution (polling), or to maintain an ongoing connection and receive live updates (streaming).
Collaboration agents would use the same part of the protocol to monitor the progress of the solution and supervision agents.

\item[\reqobserve: \reqobservedesc] For this experiment, we will evaluate integrating the agents with existing observability platforms for LLM-based applications, which can be run on-premises, such as Arize Phoenix~\cite{arize_phoenix} and Langfuse~\cite{langfuse_website}. Their instrumentations send OpenTelemetry~\cite{opentelemetry_website} traces to their servers. To unify these traces across a collaboration involving a tree of tasks and sub-tasks to be solved, the protocol would need to support \emph{super-task IDs} that relate agent traces to higher-level tasks.


\item[\reqinterop: \reqinteropdesc] The protocol should not arbitrarily constrain developers' choices with respect to implementation technologies, such as programming languages or agentic frameworks. We specifically selected a software engineering task that necessitates the use of multiple programming languages. Since the Emfatic parser is implemented in Java, the syntactic supervision agent must be implemented in Java. On the other hand, many contemporary agentic frameworks are predominantly based on Python, which implies that other agents in the collaboration can naturally be implemented in Python.

\item[\reqaccess: \reqaccessdesc] Access control will be evaluated through examination of the protocol itself, rather than through the implemented agents.
It is envisioned that HTTP-based protocols can largely treat authentication in an \emph{off-band} manner, through existing HTTP mechanisms (e.g., additional headers or TLS encryption).
Authorization will require having some way to signal the refusal to comply with a specific request, or to provide access to a specific agent.

\end{description}


\section{Implementation}

This section details the implementation of the use case described in Section~\ref{sec:task} using MCP and A2A. Thus, two equivalent implementations have been realised: one based on MCP (see Figure~\ref{fig:mcp_architecture}) and another based on A2A (see Figure~\ref{fig:a2a_architecture}).

\begin{figure}[t]
    \centering
    \includegraphics[width=0.99\linewidth]{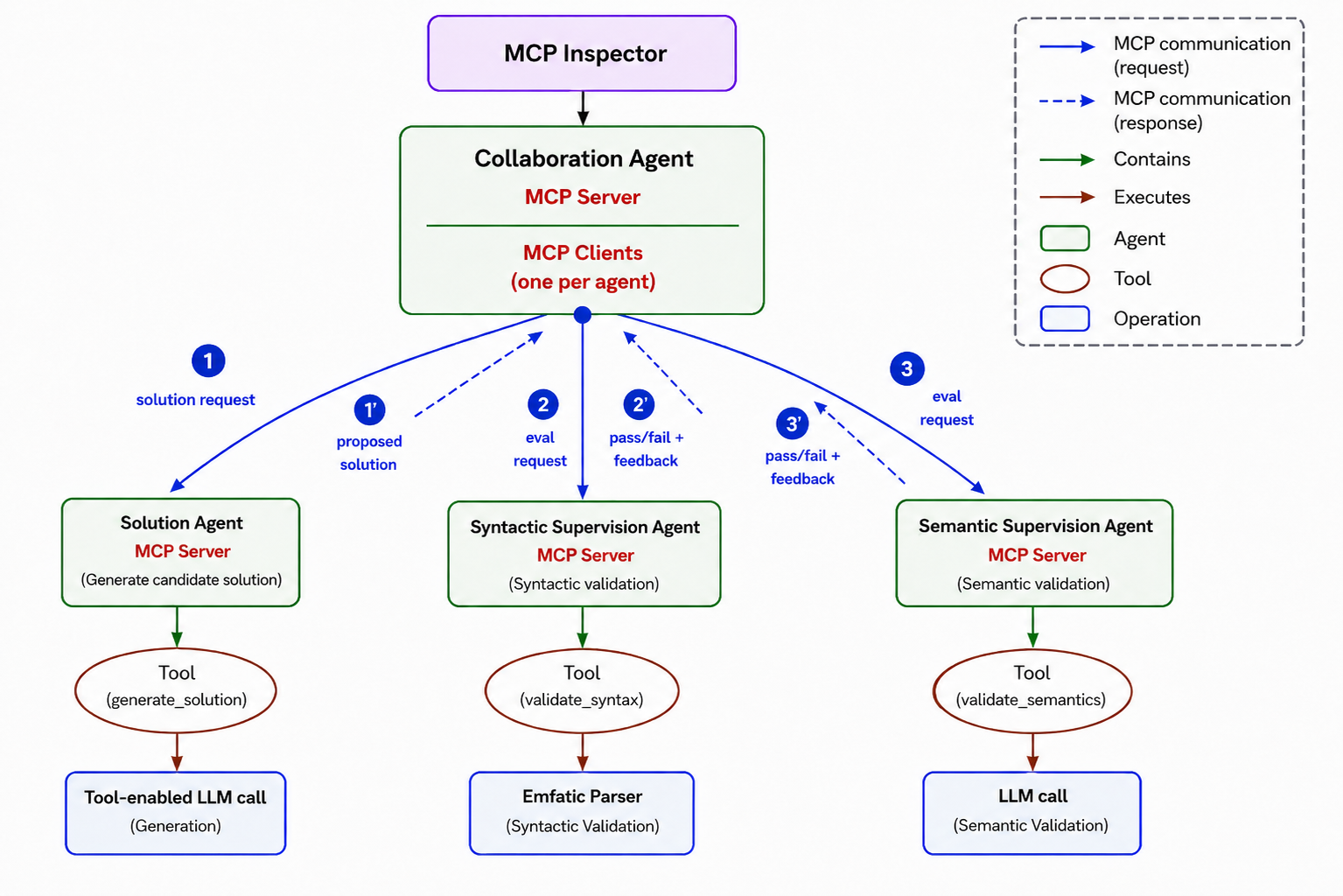}
    \caption{MCP-based system architecture}
    \label{fig:mcp_architecture}
\end{figure}

\subsection{MCP-Based Implementation}
\label{sec:mcp-implementation}

In the MCP multi-agent system implementation, each agent is exposed as an MCP server that contains one tool, as depicted in Figure~\ref{fig:mcp_architecture}. The tool of an agent's MCP server carries out the actions of the agent; for example, the solution agent's MCP server exposes a tool that calls an LLM for proposing a solution. However, the MCP tool encapsulating the solution agent's behaviour also provides the LLM with additional tools for building the Emfatic source code by following the ReAct architecture, as detailed in Section~\ref{sec:task}. In this setting, the tool encapsulating the agent's behaviour corresponds to the root-level tool, in which other tools are called. Therefore, although tools and agents are not strictly identical concepts, the root-level tool becomes nearly equivalent to the agent in this setting, as it effectively encapsulates the behaviour and execution logic of the agent. The user uses an MCP client (e.g., the MCP Inspector \cite{mcp-inspector}) to connect to the MCP server associated with the collaboration agent, and then calls its tool with a prompt as a parameter. Then, the collaboration agent uses an MCP client to connect to the MCP server of the solution agent, and calls the tool of the solution agent by passing on the prompt as a parameter. The solution agent computes the solution using an LLM that leverages other internal tools, and returns the proposed solution to the collaboration agent. The remainder of the workflow is the same as in Section~\ref{sec:task}, with the difference that inter-agent communication entails an MCP client connecting to an MCP server for calling a tool wrapping the agent's behaviour.

The source code of the implemented MCP-based system, written in Java and Python and relying on MCP SDK APIs~\cite{mcp_sdk}, is available in~\cite{implementations_mcp_a2a}. The collaboration and syntactic supervision agents do not use LLMs. The solution and semantic supervision agents use the qwen2.5-coder:32b LLM.

The solution agent, the syntactic supervision agent, and the semantic supervision agent send only the final solution to the collaboration agent. The collaboration agent sends to the calling MCP client logging notifications regarding agent invocations, including input and output messages. Moreover, Langfuse was configured to enable observability, to trace the activity of the collaboration.


An alternative implementation based on MCP would avoid direct agent invocation and instead rely on shared communication artifacts, such as mailboxes or task queues, exposed as MCP resources or tools. In this approach, agents would coordinate indirectly by publishing to and consuming from shared resources, with MCP providing the underlying integration infrastructure. This design closely aligns with MCP's original emphasis on structured access to external resources and could support coordination patterns such as asynchronous messaging and persistent shared state. However, adopting this approach would require conversation management, message correlation, and execution context management to be implemented at the application level rather than being supported by the protocol itself. For this study, we therefore focus on evaluating MCP under its official specification and intended usage model, in order to ensure both a clearer and a fairer comparison with A2A's protocol-level support for inter-agent communication, while considering infrastructure-mediated coordination as a direction for future work.


\subsection{A2A-Based Implementation}

The source code of the implemented A2A-based system, written in Java and Python and relying on A2A SDK APIs~\cite{a2a_sdk}, is available in~\cite{implementations_mcp_a2a}. The architecture of the A2A-based system is illustrated in Figure~\ref{fig:a2a_architecture}. The collaboration and syntactic supervision agents do not use LLMs. The solution agent and semantic supervision agent use the qwen2.5-coder:32b LLM. To support monitoring and debugging, observability is enabled through Langfuse, which provides detailed logs and traces of agent interactions.

The implementation leverages the streaming and task lifecycle management capabilities in A2A to orchestrate communication and manage task execution between agents. When a user sends a prompt, the collaboration agent creates a task with a task ID and context ID, which serve as persistent identifiers throughout the entire workflow. Each agent can update the task by generating a \classname{Task\-Status\-Update\-Event} object containing state information with \classname{TaskStatus} objects, and human-readable messages with the A2A \classname{Message} objects. The artifact management system utilises \classname{Task\-Artifact\-Update\-Event} objects to send results between agents, with each artifact containing metadata and actual content in multi-part objects. The solution agent produces artifacts containing the generated Emfatic model, and the supervisor agents generate artifacts with validity statuses and feedback. The collaboration agent acts as an orchestrator, and sends status updates at key milestones: agent discovery, agent invocations, result update, retry attempts, and final completion or failure. 

\begin{figure}[t!]
    \centering
    \includegraphics[width=0.99\linewidth]{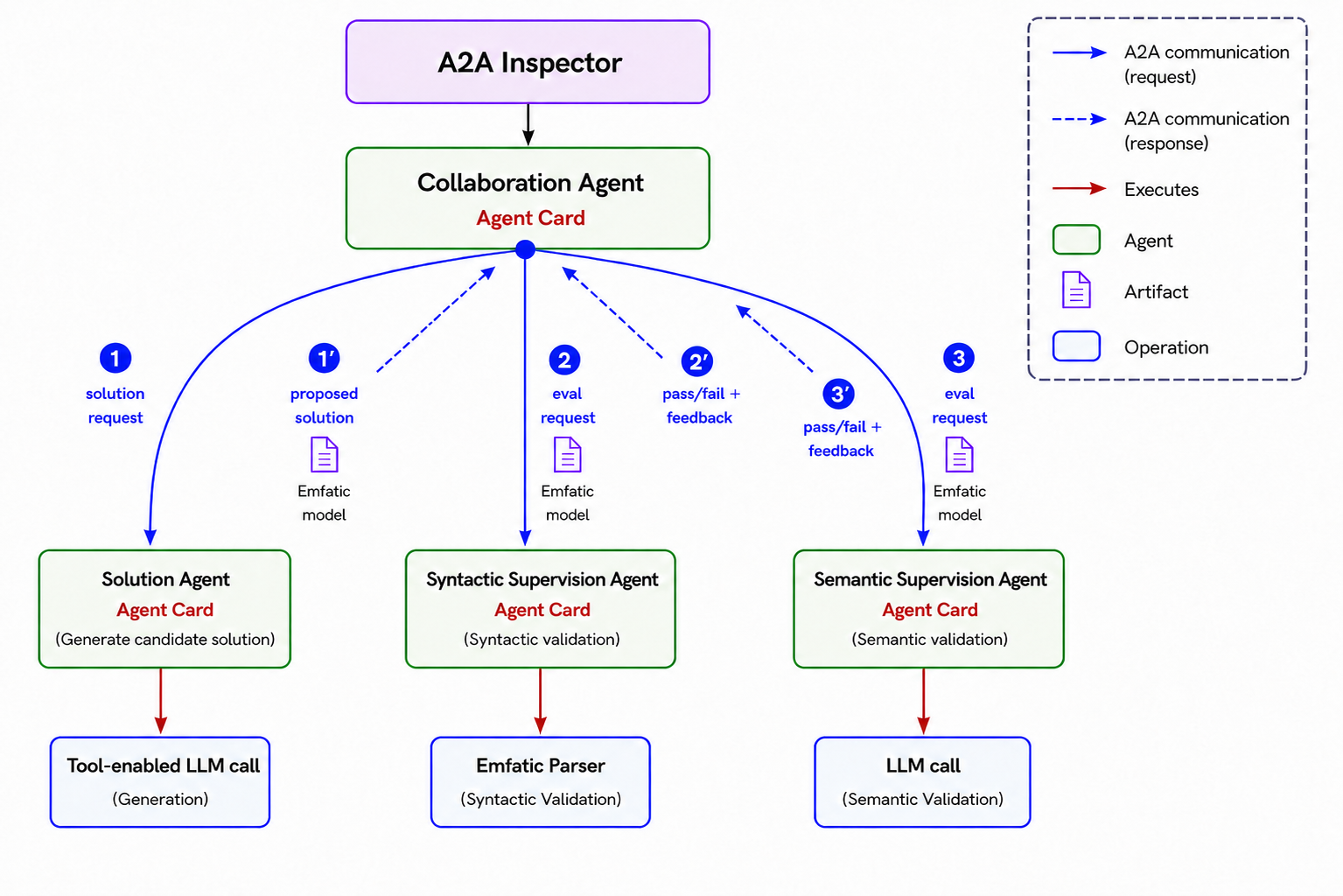}
    \caption{A2A-based system architecture}
    \label{fig:a2a_architecture}
\end{figure}

The collaboration workflow implements retry logic with configurable maximum attempts, where validation failures trigger regeneration cycles with enriched context from supervisor feedback. Throughout this process, the task ID and context ID remain constant, ensuring correlation of messages and artifacts, and enabling the reconstruction of the complete interaction history. The final successful completion is signalled through a \classname{Task\-Status\-Update\-Event} with state set to completed and final set to true, accompanied by consolidated artifacts containing the validated Emfatic code and comprehensive validation and feedback from all participating agents. 

\section{Evaluation}

The MCP-based and A2A-based implementations were validated using 30 prompts, of which 10 were straightforward and 20 were of higher complexity. The validation prompts are available in~\cite{implementations_mcp_a2a}. Listing~\ref{lst:user-prompt} contains one of the validation prompts, which requests the generation of a library model. Listing~\ref{lst:mcp-generated-model} presents the solution produced by the solution agent in response to the prompt in Listing~\ref{lst:user-prompt}, which was successfully validated both syntactically and semantically.

\begin{figure}[t]
\centering
\begin{minipage}[t]{0.53\textwidth}
\begin{lstlisting}[basicstyle=\scriptsize\ttfamily,captionpos=b,breaklines=true,breakatwhitespace=true,frame=lines,belowcaptionskip=-10pt,breakindent=0pt,caption={User prompt for a library model provided to the collaboration agent},label=lst:user-prompt]
Please generate a model of a simple library management system with three main concepts: Library, Book, and Author. Each Library has a name (String) and contains a collection of Books and a collection of Authors (both containment references). Each Book has a title (String) and a page count (int) and has a non-containment reference to its main Author. Each Author has a name (String) and age (int) and knows which Books they have written (non-containment reference to Books). The system should clearly distinguish between containment and non-containment relationships: A Library owns its Books and its Authors (deleting the Library deletes its Books and its Authors). Authors and Books reference each other but do not manage each other's lifecycles.
\end{lstlisting}
\end{minipage}
\hfill
\begin{minipage}[t]{0.4\textwidth}
\begin{lstlisting}[language=EmfaticText,basicstyle=\scriptsize\ttfamily,caption={Emfatic model generated by the solution agent},belowcaptionskip=-10pt,label=lst:mcp-generated-model]
package LibraryManagement;

class Model {
  val Library[*] libraries;
}
class Library {
  attr String name;
  val Book[*] books;
  val Author[*] authors;
}
class Book {
  attr String title;
  attr int pageCount;
  ref Author author;
}
class Author {
  attr String name;
  attr int age;
  ref Book[*] books;
}
\end{lstlisting}
\end{minipage}
\end{figure}

\subsection{Analysis of Implementation and Coordination Complexity}

Table~\ref{tab:implementation_comparison} presents a comparison between the MCP-based and A2A-based implementations in terms of implementation and coordination complexity. The A2A-based implementation contains a larger codebase, comprising 1898 lines of code compared to 1255 lines for the MCP-based implementation, and thus requiring approximately 51\% more lines of code. Note that these figures were obtained by analysing the Python and Java code, excluding configuration files, using cloc~\cite{cloc_repo}, a language-aware line counting tool that automatically analyses source code repositories and reports the number of source lines of code per programming language, excluding blank lines and comments.

The A2A-based implementation employs a larger number of coordination primitives, using 10 coordination-related abstractions compared to 4 in the MCP-based implementation. In particular, the A2A-based implementation relies on task management, event propagation, structured messaging, and streaming-related primitives, whereas the MCP-based implementation relies on transport initialisation, tool invocation, and logging coordination primitives. Furthermore, the A2A-based implementation traverses a larger number of coordination stages per interaction, requiring 6 coordination stages per interaction compared to 2 in the MCP-based implementation. Note that a coordination stage corresponds to one distinct coordination step required to complete an interaction, where an interaction represents one complete delegation cycle between two agents. The MCP-based implementation involves two stages: transport initialisation and tool invocation. Conversely, the A2A-based implementation involves several stages: agent discovery, task creation, message dispatch, task state progression, artifact propagation, and completion handling. This reflects the richer protocol-level workflow semantics and lifecycle management capabilities provided by A2A. Consequently, the comparatively smaller number of coordination primitives and stages per interaction enables the MCP-based implementation to follow a lighter-weight coordination model with lower orchestration complexity and a smaller coordination failure surface for inter-agent communication workflows.

\begin{table}[t]
 \caption{Comparison of implementation and coordination characteristics}
  \label{tab:implementation_comparison}
  \centering
  {\fontsize{8}{9}\selectfont
  \begin{tabular}{l c c c}
    \toprule
    Metric & MCP & A2A \\
    \midrule
    Lines of code & 1255 & 1898 \\
    Coordination primitives & 4 & 10 \\
    Coordination stages per interaction & 2 & 6 \\
    \bottomrule
  \end{tabular}
  }
 \vspace{-1mm}
\end{table}

\subsection{Analysis of Requirements Fulfilment}

The MCP-based and A2A-based implementations have been evaluated against the requirements introduced in Section~\ref{sec:requirements-evaluation}. As detailed in Table~\ref{tab:req-eval-summary}, the MCP protocol is evaluated as meeting 5 requirements (\reqdiscover, \reqmultipart, \reqasync, \reqinterop, \reqaccess), partially meeting 1 requirement (\reqobserve), and not meeting 1 requirement (\reqmultiturn). Conversely, the A2A protocol is evaluated as meeting 6 requirements (\reqdiscover, \reqmultipart, \reqmultiturn, \reqasync, \reqinterop, \reqaccess), and partially meeting 1 requirement (\reqobserve).

\begin{description}

\item[\reqdiscover: \reqdiscoverdesc] MCP enables discoverability through MCP servers exposing tools and capabilities that can be queried by clients, whereas A2A provides explicit discoverability through Agent Cards and Agent Skills.

\item[\reqmultipart: \reqmultipartdesc] MCP tools support structured input and output schemas containing multiple logical parts, whereas A2A messages support different \classname{Part} types (e.g., text, file, data). Structured information is exchanged between agents without requiring ad-hoc serialisation.

\item[\reqmultiturn: \reqmultiturndesc] A2A supports iterative interactions and clarification requests through task lifecycle states such as \emph{input-required}. MCP does not provide native task or conversation tracking primitives, requiring servers to explicitly implement their own context tracking mechanisms.

\item[\reqasync: \reqasyncdesc] MCP supports asynchronous request/response handling, event notifications, and streaming through Server-Sent Events (SSE) and Streamable HTTP, whereas A2A supports streaming and asynchronous operations through SSE and push notifications.

\item[\reqobserve: \reqobservedesc] Neither protocol provides native protocol-level tracing or explicit \emph{super-task IDs} for observing an entire multi-agent collaboration. However, both implementations integrate external observability mechanisms through callback functions or third-party libraries.

\item[\reqinterop: \reqinteropdesc] MCP provides transport-agnostic interoperability through standardised protocol messages and official SDKs, whereas A2A provides interoperability through HTTP-based communication and multiple SDKs.

\item[\reqaccess: \reqaccessdesc] MCP supports access control via server-side mechanisms~\cite{mcp_authorization} and SDK APIs, whereas A2A integrates authentication and authorization through standard web mechanisms such as OAuth2 and OpenID Connect.

\end{description}

\newcommand*{\reqmet}{$\checkmark$}
\newcommand*{\requnmet}{$\times{}$ }
\newcommand*{\reqpartial}{$\sim{}$}

\begin{table}[t]
 \caption{Requirement evaluation summary: \reqmet is met, \requnmet is not met, \reqpartial is partially met}
  \label{tab:req-eval-summary}
  \centering
  {\fontsize{8}{9}\selectfont
  \begin{tabular}{l c c c}
    \toprule
    Requirement & MCP & A2A \\
    \midrule
    \reqdiscover: \reqdiscoverdesc & \reqmet & \reqmet \\
    \reqmultipart: \reqmultipartdesc & \reqmet & \reqmet \\
    \reqmultiturn: \reqmultiturndesc & \requnmet & \reqmet \\
    \reqasync: \reqasyncdesc & \reqmet & \reqmet \\
    \reqobserve: \reqobservedesc & \reqpartial & \reqpartial \\
    \reqinterop: \reqinteropdesc & \reqmet & \reqmet \\
    \reqaccess: \reqaccessdesc & \reqmet & \reqmet \\
    \bottomrule
  \end{tabular}
  }
 \vspace{-1mm}
\end{table}

The results indicate that the MCP-based implementation supports agent discoverability, multi-part messaging, asynchronous and streaming communication, interoperability, and access control, while it does not provide native support for multi-turn conversations and only partially supports agent observability. In comparison with the A2A-based implementation, MCP satisfies the requirements in largely the same way, with the notable exception of multi-turn conversations, which are explicitly supported by A2A but not natively supported by MCP, while observability remains only partially supported in both. The key distinction with regard to these requirements concerns multi-turn, stateful task handling: A2A provides explicit protocol-level support for long-running tasks, task states, and \emph{input-required} interactions, whereas MCP requires these aspects to be realised through explicit orchestration logic implemented in the application layer.


\section{Conclusions}

This paper presented an empirical report on the practical use of MCP and A2A, by implementing a multi-agent collaboration that solves a common software engineering task: given a natural language prompt, generate an object-oriented domain model. The results evidence that MCP can support inter-agent coordination in constrained LLM-based systems through a comparatively lightweight implementation model with lower coordination complexity, although coordination concerns such as conversational state management and task lifecycle handling must be implemented explicitly at the application layer. Conversely, A2A provides richer native support for stateful, multi-turn coordination through protocol-level abstractions for tasks and lifecycle management, but this comes with substantially greater implementation and coordination complexity. Considering the narrow scope of the evaluated coordination pattern, these findings are intended as design observations rather than general claims of protocol suitability across broader classes of MAS. As future work, we plan to investigate dynamic, LLM-driven inter-agent coordination, in which coordination structures are synthesised or adapted at runtime rather than fixed a priori.

\begin{credits}
\subsubsection{\ackname} This paper was supported by the MOSAICO project (Management, Orchestration and Supervision of AI-agent COmmunities for reliable AI in software engineering) funded by the EU Horizon Research and Innovation Action (Grant Agreement No. 101189664).
\end{credits}

\bibliographystyle{splncs04} 
\bibliography{mosaico}

\end{document}